\documentclass[pra,amsmath,amssymb,,showpacs]{revtex4}

\usepackage{epsfig}

\usepackage{color}
\usepackage{ulem}

\begin{document}

\title{Maximally Entangled States}

\author{M. Revzen}
\affiliation {Department of Physics, Technion - Israel Institute of Technology, Haifa 32000, Israel}

\date{\today}

\begin{abstract}

Every Maximally Entangled State (MES) of two d-dimensional particles is shown to be a
product state of suitably chosen collective coordinates. The state may be viewed as
defining a "point" in a "phase space" like $d^2$ array representing  $d^2$ orthonormal
Maximally Entangled States  basis
for the Hilbert space.\\
A finite geometry view of MES is presented and its relation with the afore mentioned "phase
space" is outlined: "straight lines" in the space depict product of single particle
mutually unbiased basis (MUB) states, inverting thereby Schmidt's diagonalization scheme
in giving a product
single particle states as a d-terms sum of maximally entangled states. \\
To assure self sufficiency the essential mathematical results are summarized in the
appendices.
\end{abstract}


\maketitle

\section{Introduction}

The most remarkable attribute of quantum mechanics (QM) is its association of physical
processes with linear relations among  probability {\it amplitudes}. A striking consequence
of this is the appearance of entangled states \cite{schrodinger}. These states, introduced
by Einstein, Podolsky and Rosen (EPR) \cite{epr}, were shown to be intimately involved in
almost all
subtle characteristics of QM which even today defies comfortable understanding \cite {laloe}.\\

The almost complete dominance of these states in studies of the foundation of QM studies
was instigated by Bohm who formulated the EPR considerations within a finite dimensional
Hilbert space: he considered spin states and orientations rather than position and momenta
which have continuous values. Bell's analysis \cite{bell, shimony} of the EPR - Bohm study
revealed a non locality facet that may  reside in entangled particle's pair correlation.
Introduction of a basis of orthonormal maximally entangled states \cite{sam} in this
context encouraged the formulation of protocols for teleportation \cite{asher}, entangling
QM with information theory which is, arguably, the
central theme of present day investigations of QM foundations.\\

In this work we study maximally entangled states (MES) of two d dimensional particles
(systems). Our study is confined to d a prime $\ne 2$  since for these dimensionalities the
mathematics is simplest - thus though extensions to d all primes and power of primes is
possible \cite{wootters,durt,v,klimov} it requires a more sophisticated mathematics while,
we judge, not enhancing physical transparency. We begin with a brief review of an
information theoretic definition \cite{barnett} of Maximally Entangled State (MES) (Section
II) \cite{fivel}. In Section III we show that {\it every} MES may be viewed as a product
state in appropriately selected collective coordinates. Thus, whereas the strong
correlation between the entangled particles is studied extensively, the present study
consider the independence between "collective" variables implied thereby. Such a
representation of MES as a product state (in the collective variables)  is used to
conveniently define an orthonormal MES basis for the $d^2$ dimensional Hilbert space. These
are product states of the two particles/systems "center of mass" like and "relative"
coordinates: the state is specified with phase space like parameters with the $d^2$ phase
space like points parametrizing the $d^2$ dimensional orthonormal MES basis of the whole
Hilbert space. The next section, Section IV, contains (finite) geometry notions \cite{mary,r1}for the
$d^2$ square array of phase space like points: We consider states formed by sums of states
underpinned with points (q,p) of the phase space (which are MES) that lie on a straight
line. We show that the the resultant states are (product of) single particle mutually
unbiased basis (MUB) states, revealing, thereby, a novel scheme for the construction of MUB
states. Furthermore, as is given
in Section V, it discerns inversion of the Schmidt \cite{peres,schmidt} diagonal form of MES.\\

We add four appendices. The first, Appendix A, reviews \cite{r1} the definition of "tilde"
states used in the discussion of the "universal" form of MES in Section II. Appendix B
elaborates state relabeling issue which allows close correspondence between the labeling of
the isomorphic Hilbert spaces of the two particles. Appendix C details our definition of
collective coordinates \cite{r2,r3} and indicates the source of the nomenclature "center of
mass" and "relative" coordinates, while Appendix C provides a brief
review for the particular Mutual Unbiased Bases (MUB) used in the text.\\

\section {Maximally Entangled States (MES) and Line States}

Information theoretic approach to entanglement is considered in \cite{barnett}. The present
study deals with special case thereof: maximal entanglement of two equi-dimensional (d)
particles in a pure state and their MES bases. A complete orthonormal basis, for d=2,  of maximally
entangled
states (MES) was introduced in \cite{sam}; an in-depth study of these bases is due to Fivel
\cite{fivel}. Further works dealing with this were published much later:
\cite{ban,klimov,r1}. We shall follow \cite{fivel,r1} in the main.\\

We define MES in close analogy with \cite{fivel} (This definition is, in essence, a
special case of the one given by \cite{barnett}), \\

Definition: \\

 MES of two d - dimensional particles (or systems - we
use the terms interchangeably) is a pure state, $|u>$, implying that the probability of
having particle i (i=1,2) in a arbitrary  single particle state $|\alpha>$ is independent
of i and $\alpha$.\\

It is a mathematically proven result that every two particles state is expressible in
Schmidt decomposed \cite{peres,schmidt} sum of mutually orthogonal terms each involves
product state of the two particles. One notes that the Schmidt decomposition of MES is, of
necessity, of the form

\begin{equation}\label{u}
|u(1,2)\rangle=\frac{1}{\sqrt d}\sum_{m=0}^{d-1}|m,b\rangle_1
|m-c,b'\rangle_2;\;\;c=0,1,...d-1.
\end{equation}
The sum, in Eq.(\ref{u}), {\it defines} two orthonormal bases, b and b' - the summation (mod[d])
include all d (orthonormal) states for either particle. This is {\it necessary} for if any member
m of either basis were missing the probability of having the relevant particle in that
state will be nil , contrary to the MES definition. Thus the expansion specifies two
orthogonal bases: one for each of the particles.\\
The form, Eq.(\ref{u}), is {\it sufficient} for a MES of two d-dimensional particles. This
may be verified upon partial tracing of such state with respect to either particle
coordinates leaves unity as the resultant density operator of its mate: Thus the
probability of either particle is in any single particle state is 1/d, , e.g.,the
probability of particle 2 is in the (single particle) state $|\alpha \rangle$

\begin{equation}\label{prob}
P(2,\alpha)=tr |u\rangle\langle u|\alpha \rangle_2\langle \alpha|=\frac{1}{d}.
\end{equation}

Thus we have shown that the definition of  MES as given above is both necessary and
sufficient to imply that every MES can be expressed in the form of Eq.(\ref{u}).\\

It is noteworthy that each MES allows a convenient definition for a MES basis for the entire two
d-dimensional particles' Hilbert space. Thus the $d^2$ states generated from such state,
\begin{equation}\label{u4}
|u^{b,b'}_{q,p}\rangle=\frac{1}{\sqrt d}\sum_{m}|m;b\rangle_1 \omega^{-mp}|m-q;b'\rangle_2,\;\;
q,p=0,1,...d-1,\; -c=d-c\;Mod[d],
\end{equation}
are, as can be verified by direct substitution, orthonormal
\begin{equation}
\langle u^{b,b'}_{q,p}| u^{b,b'}_{q',p'}\rangle=\delta_{q,q'}\delta_{p,p'},
\end{equation}
and thence from a complete orthnormal MES basis for the Hilbert space. We shall find it
convenient to refer to a  product state, e.g. $|m;b\rangle_1|n;b'\rangle_2$ as a "point" in
a square $d\times d$ array whose x (discrete) coordinate is m and the y coordinate is n.
Thus we may view the state in Eq.(\ref{u4}) as a "line" state - it runs over d "points"
(the summation is modular, mode[d]) underpinning the line
$m(n)=n+q\;Mod[d]$. Schmidt decomposition is seen thereby to be a line state expression for MES.\\

A perhaps physically more transparent \cite{fivel} form for arbitrary  MES (e.g. our
example, Eq.(\ref{u}), obtains upon relabelling
$|m-q;b'\rangle\rightarrow|\tilde{m};\tilde{b}\rangle$ (cf. Appendices A,B). The state in
this case is the universal (i.e. basis independent, \cite{r1,r2}), state
$|{\cal{R}}\rangle$:
\begin{equation}
|u(1,2)\rangle \rightarrow |{\cal{R}}\rangle=\frac{1}{\sqrt d}
\sum_m|m;b\rangle_1|\tilde{m};\tilde{b}\rangle_2.
\end{equation}

\section {Maximally Entangled States in Collective Coordinates}

The collective coordinates, defined in Appendix B, are referred to as "center of mass", c,
and relative, r. Each relates, of course, to its corresponding state enumerator operator, Z
and state shifting operator, X. The eigenstates of X are the Fourier transforms of those of
Z. The latter's eigenstates are referred as the CB for the respective collective
coordinate: In the following the eigenvalues of Z (for all systems: particles 1, 2 and
collective coordinates c,r) will be denoted by q and the basis, b,  by $\ddot{0}$, those of
X by p and the corresponding basis, b, is designated by b=0; q,p=0,1,...d-1. Thus
$$Z_s|q;\ddot{0}\rangle_s =\omega^q|q;\ddot{0}\rangle_s;\;\;s=1,2,c,r;\;\;\langle q;\ddot{0}|p;0\rangle
=\frac{\omega^{-qp}}{\sqrt{d}}.$$

As noted in Appendix B, relabeling is a bas is operation. Thus relabeling
$|m-q;b'\rangle_2\rightarrow|m-q;b\rangle_2$ gives for the arbitrarily chosen MES,
Eq.(\ref{u4}),
\begin{equation}
|u^{b,b'}_{2q,p}\rangle=\frac{1}{\sqrt d}\sum_m |m;b\rangle_1\omega^{-mp}|m-2q;b'\rangle\rightarrow
\frac{1}{\sqrt d}\sum_m |m;b\rangle_1\omega^{-mp}|m-2q;b\rangle\rightarrow\frac{1}{\sqrt d}\sum_m
|m;\ddot{0}\rangle_1\omega^{-mp}|m-2q;\ddot{0}\rangle,
\end{equation}
where the line state given at the RHS is in the computational basis (designated by $\ddot{0}$) for both
particles which is most convenient for the transformation to the collective coordinates \cite{r1,r2}, and
Appendix C. Thus recalling \cite{r1,r2} and appendix C the transformation is affected by
\begin{equation}
|m;\ddot{0}\rangle_1|m';\ddot{0}\rangle_2=
|\frac{m+m'}{2};\ddot{0}\rangle_c |\frac{m-m'}{2};\ddot{0}\rangle_r.
\end{equation}
Thus
\begin{eqnarray}
|u^{b,b'}_{2q,p}\rangle &\rightarrow& \frac{1}{\sqrt d} \sum_m|m;\ddot{0}\rangle_1\omega^{-mp}|m-2q;
\ddot{0}\rangle \nonumber \\
&=&\big(\frac{1}{\sqrt d}\sum_m|m-q;\ddot{0}\rangle_c \omega^{-mp}\big)|c;\ddot{0}\rangle_r \nonumber \\
&=&\frac{\omega^{-qp}}{\sqrt d}\sum_m|m\rangle_c\omega^{-mp}|q\rangle_r=\omega^{-qp}|p;0\rangle_c
|q;\ddot{0}\rangle_r.
\end{eqnarray}
Demonstrating that every (the state considered was arbitrary) MES is a product state in collective
coordinates: in the formalism considered the "center of mass" collective coordinate has its value in
momentum space (it is the Fourier component of the CB) while the "relative" coordinate has its value in
the CB. We have thus that the two d-dimensional particles' Hilbert space is spanned by the $d^2$
orthonormal phase space like points
\begin{equation}\label{pq}
|u(p,q)+\rangle=|p;0\rangle_c|q;\ddot{0}\rangle_r=Z^{-p}_cX^q_r|0;0\rangle_c|0;\ddot{0}\rangle
;\;\;q,p=0,1,...d-1.
\end{equation}
One readily verifies that these are MES:
\begin{equation}\label{pq1}
|p;0\rangle_c|q;\ddot{0}\rangle_r=\frac{1}{\sqrt d}\sum_m|m;\ddot{0}\rangle_c\omega^{-mp}|q;\ddot{0}\rangle_r=\frac{1}{\sqrt d}\sum_m|m+q\rangle_1\omega^{-mp}|m-q\rangle.
\end{equation}

The formalism suggests the definition of a conjugate "phase space" like basis,
$|u(p,q);-\rangle$,
\begin{equation}\label{-pq}
|u(p,q);-\rangle\equiv|q;\ddot{0}\rangle_c|p;0\rangle_r
=X_c^qZ_r^{-p}|0;\ddot{0}\rangle_c|0;0\rangle_r.
\end{equation}

Thus MES have a lattice phase space structure \cite{fivel} wherein a lattice point (q,p) is
realized by collective coordinates product state
$(q,p)\Leftrightarrow|q;\ddot{0}\rangle_c|p;0\rangle_r.$ Studies of dynamically induced
"hopping" on this lattice \cite{fivel} generated by unitary operator U(t) that cause the
two particles to evolve such that at discrete times, t=0,1,2..., their state will coincide
with a lattice site is direct within the collective coordinates formulae: U(t) is a simple
product of powers of the collective coordinates operators, $X_{c},Z_{c};X_{r},Z_{r}$. e.g.
\begin{eqnarray}
X_{c}^{t=2}X_{r}^{3(t=2)}|q;\ddot{0}\rangle_c|p;0\rangle_r&=&\omega^{6p}|q+2;\ddot{0}\rangle_c|p;0\rangle_r
\nonumber \\
(q,p)&\Rightarrow& \omega^{6p}(q+2,p).
\end{eqnarray}
We now note a remarkable attribute of MES, striking especially in cases where the particles
are widely separated: Let us act on one of the particles, e.g., on particle 1, with
$\hat{X}^2_1$. The state of the particle (1 in our case) is unaffected: the probability of
it being in any single particle $|\alpha\rangle$ remains 1/d - independent of $\alpha$.
However the collective coordinates are affected, will undergo a simple change in this
example. Explicitly,
\begin{eqnarray}
\hat{X}^{2}_{1}\frac{1}{\sqrt d}\sum_{m}|m;b\rangle_1|\tilde{m};\tilde{b}\rangle_2&=&\frac{1}{\sqrt d}
\sum_{m'}|m';b\rangle_1\sum_{m}U^{b}_{m',m}|\tilde{m};\tilde{b}\rangle_2= \nonumber \\
\frac{1}{\sqrt d}\sum_{m'}|m';b\rangle_1 \sum_{m} U^{\tilde b}_{\tilde{m}{'},\tilde{m}}
|\tilde{m};\tilde{b}\rangle&=&\frac{1}{\sqrt d}
\sum_{m'}|m';b\rangle_1|\tilde{m}{'};\tilde{b}\rangle,
\end{eqnarray}
i.e. the state remains a MES where either particle may be found with equal probability in any single
particle state, i.e. neither particle's expectations values of any single particle operator is affected.\\

We now consider the effect on the collective coordinates state. The universal state
representation of the MES, was shown above to imply
\begin{equation}
\frac{1}{\sqrt
d}\sum_{m}|m;b\rangle_1|\tilde{m};\tilde{b}\rangle_2=|0;0\rangle_c|0;\ddot{0}\rangle_r.
\end{equation}
Thus,
\begin{eqnarray}
\hat{X}^2_1\frac{1}{\sqrt
d}\sum_{m}|m;b\rangle_1|\tilde{m};\tilde{b}\rangle_2&=&\hat{X}_r\hat{X}_c|0;0\rangle_c|0;\ddot{0}\rangle_r
=\nonumber \\
\hat{X}_r\hat{X}_c\frac{1}{\sqrt
d}\sum_n|0;0\rangle_c|n;0\rangle_r&=&|0;0\rangle_c|-1;\ddot{0}\rangle_r,
\end{eqnarray}
i.e. the states collective coordinates is affected.

\section{Mutually Unbiased Bases and the Inversion of Schmidt decomposition}

Mutual unbiased bases (MUB) were introduced by Schwinger \cite{schwinger} as the bases
associated with operators of "maximum degree of incompatibility", the information theoretic
oriented appellation "Mutually Unbiased Bases" (MUB) that is essentially universal now, was
dubbed by Wootters \cite{wootters}. Appendix D gives a brief theoretical overview of MUB.
MUB, aside from their direct theoretical significance as representatives of the uncertainly
among (relevant) observables, are widely used in studies of foundations of quantum
mechanics (QM), information and cryptography. Several approaches to their construction are
known \cite{wootters,gib,v,tal,durt,klimov,r1},
 we present here a new scheme,
underscoring their relation to MES \cite{saniga,r1} which may provide new insight on their role in QM.\\

We have shown above, Eq.(\ref{pq}), that an arbitrary MES $|u\rangle$, defines and thus
provides an "origin" for $d^2$ orthonormal states, $u(q,p)\rangle =
|q;\ddot{0}\rangle_c|p;0\rangle_r\;q,p=0,1,...d-1.$ (Recall that $\ddot{0}$ dubs the CB,
the eigenfunctions of $\bar{Z}$, while 0 (in the vectorial label, dubbed $|p;0\rangle_r$) -
eigenfunctions of $\bar{X}$.) These states now realize points in a $d^2$ square array whose
rows (lines parallel to the  x axis) are enumerated by p and its columns by q (parallel to
the y axis). This d by d square will be referred to as "phase space". We define a line, L,
in this phase space, as the aggregate of d points given by linear relation between p and q.
E.g. for d=3 the line given by the equation, p=0, consists of the three points whose
coordinates are (0,0);(1,0) and (2,0), whereas the line p=q contain the points (0,0);(1,1)
and (2,2). In our study the points are realized by states. Thus in the first case above the
three states are $|0;\ddot{0}\rangle_c|0;0\rangle_r;\;
|1;\ddot{0}\rangle_c|1;0\rangle_r;\;|2;\ddot{0}\rangle_c|2;0\rangle_r.$ We now define a
{\it line state} (yet to be normalized) by the {\it sum} of its aggregate points. Since the
line is determined by two of its points the generic designation of a line state is
 $|L[(q_0,p_0);(q_1,p_1)]\rangle$ (we shall abbreviate the notation below). There are d+1 distinct lines
 emerging  from the origin realized by
$|0;\ddot{0}\rangle|0;0\rangle$: d of these are given by $p=bq;\;Mod[d];\;b=0,1,...d-1$ and
additional line along the x axis whose line equation is q=0, is given by the points
$|q;\ddot{0}\rangle|0;0\rangle;\;q= 0,1,...d-1.$ There are d parallel lines to each of the
d+1 lines. e.g., for d=7, the lines parallel to $p=3q;\;Mod[7]$ are the d lines given by
$p=3q+s,\;Mod[7];\;p,q,s=0,1,...6$. Another set of parallel lines is $p=s;\;s=0,1,...6$. It
is obvious that {\it each} of the d+1 lines and its d parallels contain exactly all $d^2$
points. We label d of these d+1 sets by their b value (the set  parallel to q=0 by
$\ddot{0}$) i.e. a line is specified by its set b which gives its orientation and s that
specifies its position relative to the line of the same orientation that goes through the
origin.  (This labeling will not give rise to any conflict with the previous notation.) We
now assert that each of these sets is an MUB basis and each of the line state is a product
of MUB states, one for each particle. This is the novel construction of the MUB states.
We now prove the assertion.\\

Consider first the perhaps simplest case: the line q=m that runs parallel to the y axis at
x=m, i.e. the line with $b=\ddot{0}$ and s=m, this state line $|L(\ddot{0};m)\rangle$,(we
ignore the normalization) is:
\begin{equation}\label{lll}
|L(\ddot{0};m)\rangle=\sum_p |m;\ddot{0}\rangle_c|p;0\rangle_r =\sum_{p,n}
|m;\ddot{0}\rangle_c|n;
\ddot{0}\rangle_r\omega^{-np}=|m;\ddot{0}\rangle_c|0;\ddot{0}\rangle_r=
|m;\ddot{0}\rangle_1|m;\ddot{0}\rangle_2.
\end{equation}
We used $\sum_p\omega^{pn}=d\delta_{n,0}$ and $n_1=n_c+n_r;\;\;n_2=n_c-n_r.$ Thus we got
that the line state is a product state of a particle 1 state with its tilde state of
particle 2 (note for $b=\ddot{0}$
the tilde state is identical with its mate).\\

The study thus far indicated that Schmidt decomposition of MES may viewed as expressing the
MES as a "line state" made of "points" each of which is a product state in the particle
coordinates. The "line state"  and thereby the MES it account for was shown to be a product
state in appropriately chosen collective coordinates. This suggests considering product
state (in the particle coordinates) be related a "line state" in the phase space like
square array whose constituent points space (q,p) that underpin product states in
collective coordinates. Such an account of particles product state is termed inversion of
the Schmidt decomposition. We shall now show that, indeed, summing over phase space points
on a straight line, viz. (denoting the line state by $|L(m,b)\rangle$),
\begin{equation}
|L(m,,b)\rangle=\frac{1}{\sqrt{d}}\sum_q
|p(q);0\rangle_c|q;\ddot{0}\rangle_r;\;\;p(q)=bq+m;\;q,p,b,m=0,1,...d-1,
\end{equation}
gives a product state in the single particle coordinates, i.e. invert the Schmidt
diagonalization, the line state equals a product of two single particle MUB states:
\begin{eqnarray}
|L(m,,b)\rangle&=&\frac{1}{d}\sum_q\sum_n \omega^{-np}|n\rangle_c|q\rangle_r \nonumber \\
               &=&\frac{1}{d}\sum_q\sum_n \omega^{-n(bq+m)}|n+q\rangle_1|n-q\rangle_2 \nonumber \\
               &=&\frac{1}{d}\sum_{n_1}\sum_{n_2}|n_1\rangle_1|n_2\rangle_2\omega^{-\frac{n_1+n_2}{2}\big[b(n_1-n_2)+m\big]}
               \nonumber \\
               &=&\frac{1}{\sqrt d}\sum_n|n\rangle_1\omega^{\frac{b}{4}n^2-\frac{m}{2}n}\frac{1}{\sqrt d}
               \sum_n|n\rangle_2\omega^{-\frac{b}{4}n^2+\frac{m}{2}n},\nonumber \\
               &=&|\frac{m}{2};\frac{b}{4}\rangle\rangle_2|\tilde{\frac{m}{2}};\tilde{\frac{b}{4}}
               \rangle\rangle_1.
\end{eqnarray}
We used Eq.(\ref{colx},\ref{colz},\ref{rc12},\ref{pq1}); the double angular bracket signified
and MUB, cf. Appendix D:
\begin{equation}
|m;b\rangle\rangle=\frac{1}{\sqrt d}\sum_n|n\rangle\omega^{bn^2-mn}.
\end{equation}
Thus the "line state" constituting of product collective coordinates states, i.e., MES in
particles coordinates expresses product single particle state, i.e. reversing Schmidt
diagonalization. The result yielding product MUB states provides a novel way for obtaining
MUB states.

We further note that the two bases Eq. (\ref{pq},\ref{-pq}) are mutually unbiased:

\begin{equation}
|\langle u(q'p';-)|u(q,p;+)\rangle|=\frac{1}{d};\;\forall\; q,q',p,p'.
\end{equation}
This relation provided the means for determining separately the state and the
measurement-basis in \cite{a,ab,r1}.\\

\bigskip
\section{Conclusions and Remarks}
\bigskip

Maximally Entangled States (MES) for two d-dimensional particles, $d=prime\;\ne 2$, were
defined via slightly modified study by Fivel \cite{fivel} as a pure two particle state in
which the  probability of finding particle i (i=1,2) in a single particle state
$|\alpha\rangle$ is independent of i and $\alpha$. This underscores an information
theoretic, purely quantal, attribute of entanglement
noted earlier in \cite{barnett}: stronger quantum correlation among the pair entails weaker individual
specification. Thus maximum entanglement randomizes the state of its constituents. Mathematically such a
state is a pure two particles state whose Schmidt decomposition \cite{peres,schmidt} involves a d-terms sum of
product state containing, with equal weight, full bases of the two d-dimensional Hilbert spaces. As such
it has a "universal", i.e. basis independent, characterization which has a geometrical interpretation
\cite{r3}.\\
It was argued that  {\it every} MES allows natural definition for a MES basis, viz $d^2$ orthonormal MES, spanning the two d-dimensional particles Hilbert space wherein the states, forming the basis are viewed as "line states" each being a Schmidt decomposition of a MES.\\

We then showed that {\it every} MES may be expressed as a product state in appropriately
chosen collective coordinates denoted by c ("center of mass") and r ("relative"). These are
studied in terms of their Schwinger operators, $\hat{Z}_s$,  (s=c,r), whose eigenfunctions
are the reference basis (dubbed computational basis (CB) almost universally and which we
denote by $b=\ddot{0}$) and $\hat{X}_s$ which is the Schwinger shift operator whose
eigenfunctions form the Fourier transform of the CB (that is denoted with b=0). The
collective coordinates provide an economic means for parametrizing MES. Thus the product
states in the collective coordinates (which are MES in the particles coordinates) are
conveniently given in terms of phase space like variables,
$|q;\ddot{0}\rangle_r|p;0\rangle_c$ (here $\ddot{0}\; and\; 0$ denotes the bases mentioned
above while q and p relates to the respective eigenvalues, $\omega^q,\;\omega^p$ with
$\omega=e^{i\frac{2\pi}{d}}\;q,p=0,1,...d-1)$. We thus have the $d^2$ dimensional Hilbert
space imaged by the $d^2$ phase space like square whose points are specified by (q,p) which
are realized by the MES
 $|q;\ddot{0}\rangle_r|p;0\rangle_c$.\\

We demonstrated that acting on the individual particles in a maximal entangled state leaves the
particles unaffected while modifying the phase space like collective coordinates of the state.\\

We then considered "line states" $|L(m,b)\rangle$ which are defined (unnormalized) as the
sum of MES along straight lines in the phase space like coordinated with (q,p),
q,p=0,1,...d-1. These are p=bq-m Mod[d], for $b \ne \ddot{0}$ and q=m' for $b=\ddot{0}$.
Thus the parameter (m,b) with $b=\ddot{0},0,1..d-1$ and m=0,1,...d-1 define d+1 families of
parallel lines each containing d phase space points (q,p). Each family contains the whole
$d^2$ phase space points once. We showed that each line state formed by adding all the
product states (in the collective coordinates) parametrized by the phase space points (q,p)
lying on the line is a two particles product state, i.e. effecting a reversal of the
Schmidt decomposition  \cite{peres,schmidt} in giving a d-terms sum expansion of product
states in terms of maximally entangled states (MES). These single particle product states
turned out to be
 Mutual Unbiased Bases (MUB) states (reviewed in Appendix C). Thus the study led to a
novel scheme for obtaining MUB which are under intensive study in the fields of foundation
of quantum mechanics, information theory and cryptography. \\
The association of {\it every} maximally entangled state (MES) with well defined collective
coordinates underscores a remarkable attribute of MES: Acting on one of its constituent
particles, e.g. with Schwinger's shifting operator $X_1$ on particle 1 which moves the
particle from computational basis state $|n\rangle_1$ to the state $|n+1\rangle_1$, does
{\it not} change the particle state. Thus the probability of finding the particle in {\it
any} single particle state $\alpha$ {\it remains independent} of $\alpha$. However the the
two particles coordinate state does change.

\bigskip
\appendix {\bf Appendix A: The "Tilde" States}\\
\bigskip

As noted by \cite{schwinger} the d-dimensional Hilbert space is spanned by an orthonormal,
 computational basis (CB), $|n\rangle,\;\;n=0,1,...d-1$ and all physical operators are
functions of two operators, $\hat{Z}, \hat{X}$ defined by
$$|n\rangle=|n+d\rangle,\;\;\hat{Z}|n\rangle=\omega^n|n\rangle,\;\omega=
e^{i\frac{2\pi}{d}};\;\hat{X}|n\rangle=|n+1\rangle.$$ Thus we may label equivalently all d
- dimensional Hilbert spaces each relative to its CB and choose the CB as real, i.e.
$|n\rangle^{\ast}=|n\rangle$.\\

Consider an arbitrary state, m, in a basis b :$|m,b\rangle$ given in terms of its CB by
$$|m;b\rangle=\frac{1}{\sqrt{d}}\sum_{n=0}^{d+1}U_{m,n}^b|n\rangle,$$
with U a unitary operator. The state $|\tilde{m},\tilde{b}\rangle$ is given by
$$|\tilde{m};\tilde{b}\rangle=\frac{1}{\sqrt{d}}\sum_{n=0}^{d+1}(U_{m,n}^b)^{\ast}|n\rangle,$$
is termed the "tilde" state associated with the state $|m,b\rangle$ \cite{umezawa}.\\

The tilde states has the following property,
\begin{equation}
|{\cal{R}}\rangle=\sum_{m}|m;b\rangle_1|\tilde{m};\tilde{b}\rangle_2=\sum_{\alpha}|\alpha;\beta\rangle_1
|\tilde{\alpha};\tilde{\beta}\rangle \;\;\forall\; b, \beta,
\end{equation}
i.e. the state $|{\cal{R}}\rangle$ is "universal": i.e. independent of basis \cite{r1}.\\
\bigskip

\appendix{\bf Appendix B: State Relabeling}\\
\bigskip

Given states enumeration $|n\rangle,\;n=0,1,...d-1$ and the Schwinger operators, Z the
enumerating operator (with $|n\rangle$ its eigenfunctions) and X their shifting operator
$X|n\rangle=|n+1\rangle$. An arbitrary state, $|m;b'\rangle$ is specified in terms of Z,X
and  $|n\rangle$. Thus labeling the state $|m;b'\rangle$, means that it is diagonalized
(without degeneracies) by $F_{b'}(X,Z)$: $F_{b'}(X,Z)|m;b'\rangle=\lambda(m)|m;b'\rangle$.
Relabeling it  by $|m+c;b\rangle,\;c=0,1,...d-1$, viz.
$$|m;b'\rangle\rightarrow |m+c;b\rangle,\;c=0,1,...d-1$$ where the latter is diagonalized  by $F_{b}(X,Z)$
means that for transformed Schwinger operators
$\bar{X},\bar{Z}$ (and thence, transformed CB $|n\rangle$, now are the eigenfunctions of $\bar{Z}$),
\begin{equation}
F_{b'}(X,Z)|m;b'\rangle=F_b(\bar{X},\bar{Z})|m+c;b\rangle.
\end{equation}
One readily sees that with $|m;b'\rangle=U|m+c;b\rangle$,
$\bar{X}=UXU^{\dagger};\;\bar{Z}=UZU^{\dagger}$ and
\begin{equation}
U=\sum_{n}|n;b'\rangle\langle b;m+c|
\end{equation}
$|m;b'\rangle$ may be relabeled $|m+c;b\rangle$. The relabeled states relates to the transformed
operators.\\

The procedure is now illustrated for a simple case with d=3 where we wish to relabel the states of particle 2 as CB : Let the MES given as
\begin{equation}\label{A1}
|u(1,2\rangle=\frac{1}{\sqrt 3}\sum |n\rangle_1|v_n\rangle_2,
\end{equation}
with
\begin{equation}\label{v}
|v_0\rangle=\frac{1}{\sqrt2}(|0\rangle+|1\rangle);|v_1\rangle=\frac{1}{\sqrt2}(|0\rangle-|1\rangle);
|v_2\rangle=|2\rangle.
\end{equation}
 The
The desired relabeling is for the second particle state to be relabeled as CB state, i.e.
eigenfunction of $\bar{Z}$, viz
\begin{equation}
|u(1,2\rangle=\frac{1}{\sqrt 3}\sum |n\rangle_1|v_n\rangle_2\Rightarrow\frac{1}{\sqrt 3}\sum |n\rangle_1|n\rangle_2.
\end{equation}
The diagonalizing operator for the state marked for relabeling is (note: we assigned to it
the desired spectrum, \cite{schwinger1}),
\begin{equation}
\hat{F}_{b{'}}=\frac{1}{2}(|0\rangle+|1\rangle)(\langle 0|+\langle 1| +\frac{1}{2}(|0\rangle-|1\rangle)\omega(\langle 0|-\langle 1|+|2\rangle \omega^2\langle 2|.
\end{equation}
The transformation operator,
\begin{equation}
\hat{U}=\sum_{n}|v_n\rangle\langle n|.
\end{equation}
Writing these operators explicitly in the (original) CB:
\begin{equation}
\hat{F}_{b{'}}=\begin{pmatrix}\frac{1+\omega}{2}&\frac{1-\omega}{2}&0\\\frac{1-\omega}{2}&\frac{1+\omega}{2}&0\\
0&0&\omega^2\end{pmatrix}\;\hat{U}=\begin{pmatrix}\frac{1}{\sqrt 2}&\frac{1}{\sqrt 2}&0\\\frac{1}{\sqrt 2}&-\frac{1}{\sqrt 2}&0\\0&0&1\end{pmatrix}.
\end{equation}
The state $|v\rangle$ (Eq.(\ref{v})) is now relabeled an eigenfunction of $\bar{Z}$, the "new"
Schwinger enumerator operator. Thus $\hat{F}_{b{'}} \rightarrow \bar{Z}$, while
$|v\rangle \rightarrow |n\rangle$, i.e. $|v\rangle$ is relabeled $|n\rangle$.\\

Notes:\\

\noindent1. The relabeling is a basis, b, operation. Thus relabeling
\begin{equation}
|m;b'\rangle\rightarrow |m+c;b\rangle\;\Rightarrow\;|m';b'\rangle\rightarrow |m'+c;b\rangle.
\end{equation}
\noindent 2. The procedure holds for all 1-1 relation between the state to be relabeled (m
in the analysis above) and its image (m+c in the analysis above). E.g. the image state
could be the tilde state:
$$|m;b'\rangle\rightarrow |\tilde{m};\tilde{b}\rangle.$$
\bigskip

\appendix{ \bf Appendix C: Collective Coordinates and Collective Bases} \\
\bigskip

The Hilbert space is spanned by the single particle computational bases,
$|n\rangle_1|n'\rangle_2$ (the subscripts denote the particles). These are eigenfunctions
of $\hat{Z}_i$ i=1,2:
$\hat{Z}_i|n\rangle_i=\omega^{n}|n\rangle_i,\;\omega=e^{i\frac{2\pi}{d}}.$ Similarly
$\hat{X}_i|n\rangle_i=|n+1\rangle_i,\;i=1,2$. Thus the exponents are modular variables. We
now define our collective coordinates and collective operators (we remind the reader that
the exponents are modular variables, e.g. 1/2 mod[d=7]=(d+1)/2)=4):
\begin{equation}\label{colz}
\hat{Z}_r\equiv \hat{Z}^{1/2}_{1}\hat{Z}^{-1/2}_{2};\;\;\bar{Z}_c\equiv
\hat{Z}^{1/2}_{1}\hat{Z}^{1/2}_{2}\;\leftrightarrow\;\hat{Z}_1=\hat{Z}_r\hat{Z}_c;\;\;\hat{Z}_2=
\hat{Z}_r^{-1}\hat{Z}_c,
\end{equation}
and, in a similar manner,
\begin{equation}\label{colx}
\hat{X}_r\equiv\hat{X}_1\hat{X}_2^{-1};\;\hat{X}_c\equiv\hat{X}_1\hat{X}_2\leftrightarrow
\hat{X}_1=\hat{X}^{1/2}_r\hat{X}^{1/2}_c,\;\hat{X}_2=\hat{X}^{-1/2}_r\hat{X}^{1/2}_c.
\end{equation}

We note that $\hat{Z}_{s}^{d}=\hat{X}_s^{d}=1,$  and
$\hat{X}_s\hat{Z}_s=\omega\hat{Z}_s\hat{X}_s,\;s=r,c;
\;\hat{X}_s\hat{Z}_{s'}=\hat{Z}_{s'}\hat{X}_s,\;s\ne s'.$ $|n_1\rangle|n_2\rangle,$ the
eigenfunctions of $\hat{Z}_i,\;i=1,2,$
 spans the $d^2$ dimensional Hilbert space.  The sets $\hat{Z}_i,\hat{X}_i;\;i=1,2$ are
 algebraically complete in this space \cite{schwinger}, i.e. every (non trivial) operator
 is a function of these operators. The eigenfunctions of $\hat{Z}_q$ are $|n_c,n_r\rangle$ with
$\hat{Z}_c|n_c,n_r\rangle=\omega^{n_c}|n_c,n_r\rangle,\;\hat{Z}_r|n_c,n_r\rangle=
\omega^{n_r}|n_c,n_r\rangle.$ We note, e.g. \cite{schwinger}, that $|n_c,n_r\rangle$ is
equivalent to $|n_c\rangle|n_r\rangle$ when, as is the present case, the two sets,
$\hat{Z}_q,\hat{X}_q;\;q=c,r$ are compatible.\\

Clearly $|n\rangle_r|n'\rangle_c;\;n,n'=0,1,..d-1,$ is a $d^2$ orthonormal basis spanning
the two d-dimensional particles Hilbert space. We may consider their respective computation
eigen bases (CB) and Fourier transform bases. To conform with previous notations
\cite{r1,r2,a,ab} the CB is designated with $\ddot{0}$ while the eigenfunction of $\bar{X}$
(the Fourier transform of   $\ddot{0}$) is designated with 0:

\begin{equation}
\hat{Z}_{s}|n;\ddot{0}\rangle_s = \omega^{n}|n;\ddot{0}\rangle_s,\;\; \hat{X}_s|n;0\rangle=
\omega^n|n;0\rangle;\;\;\langle n_s;\ddot{0}|m_s;0\rangle=\omega^{-m_sn_s}.\;s=r,c.
\end{equation}
Note: we generally adopt the abbreviation $|n;\ddot{0}\rangle\rightarrow |n\rangle$.

States in the particle coordinates may, clearly, be expressed in terms of the product
states of the collective coordinates as both form a complete orthonormal basis that spans
the two particles Hilbert space,
\begin{equation}
|n_1\rangle|n_2\rangle=\sum_{n_c,n_r}|n_c,n_r\rangle\langle n_c, n_r|n_1\rangle|n_2\rangle.
\end{equation}
The matrix element $\langle n_c, n_r|n_1\rangle|n_2\rangle$ is readily evaluated \cite{r1},

\begin{equation}\label{relcomdel}
\langle n_1,n_2|n_r,n_c\rangle=\delta_{n_r,(n_1-n_2)/2}\delta_{n_c,(n_1+n_2)/2}.
\end{equation}

Thus,
\begin{eqnarray}\label{rc12}
|n_c\rangle_c|n_r\rangle_r&=&|n_c+n_r\rangle_1|n_c-n_r\rangle_2 \nonumber \\
|n_1\rangle_1|n_2\rangle_2&=&|\frac{n_1+n_2}{2}\rangle_c|\frac{n_1-n_2}{2}\rangle_r.
\end{eqnarray}

We have then,
\begin{equation}\label{relcom}
|n_r,n_c\rangle\sim|n_1,n_2\rangle,\;\;for\;n_r=(n_1-n_2)/2,\;n_c=(n_1+n_2)/2\;\rightleftarrows
n_1=n_r+n_c,\;n_2=n_c-n_r.
\end{equation}
 There are,
of course, d+1 MUB bases for each of the collective modes. Here too, we adopt the
notational simplification $b_s\rightarrow \ddot{0}_s,\; s=r,c$.\\
\bigskip

\appendix{\bf Appendix D: Finite dimensional Mutual Unbiased Bases, MUB, Brief Review}\\
\bigskip

In a finite, d-dimensional, Hilbert space two complete, orthonormal vectorial bases, ${\cal
B}_1,\;{\cal B}_2$,
 are said to be MUB if and only if (${\cal B}_1\ne {\cal B}_2)$

\begin{equation}
\forall |u\rangle,\;|v \rangle\; \epsilon \;{\cal B}_1,\;{\cal B}_2 \;resp.,\;\;|\langle
u|v\rangle|=1/\sqrt{d}.
\end{equation}
The physical meaning of this is that knowledge that a system is in a particular state in
one basis implies complete ignorance of its state in the other basis.\\
Ivanovich \cite{ivanovich} proved that there are at most d+1 MUB, pairwise, in a
d-dimensional Hilbert space and gave explicit formulae for the d+1 bases in the case of d=p
(prime number). Wootters and Fields \cite{wootters} constructed such d+1 bases for $d=p^m$
with m an integer. Variety of methods for construction of the d+1 bases for $d=p^m$ are now
available
\cite{tal,klimov,v}. Our present study is confined to $d=p\;\ne2$.\\
 We now give explicitly the MUB states in conjunction with the algebraically complete
 operators \cite{schwinger,a} set:
 $\hat{Z},\hat{X}$.  Thus we label the d distinct states spanning the Hilbert space,
 termed
 the computational basis, by $|n\rangle,\;\;n=0,1,..d-1; |n+d\rangle=|n\rangle$
\begin{equation}
\hat{Z}|n\rangle=\omega^{n}|n\rangle;\;\hat{X}|n\rangle=|n+1\rangle,\;\omega=e^{i2\pi/d}.
\end{equation}
The d states in each of the d+1 MUB bases \cite{tal,amir}are the states of computational
basis and
\begin{equation} \label{mxel}
|m;b\rangle=\frac{1}{\sqrt d}\sum_0^{d-1}\omega^{bn^2-nm}|n\rangle;\;\;b,m=0,1,..d-1.
\end{equation}
Here the d sets labeled by b are the bases and the m labels the states within a basis.
\begin{equation}\label{tal1}
\omega^{b}\hat{X}\hat{Z}^{2b}|m;b\rangle=\omega^m|m;b\rangle.
\end{equation}
For later reference we refer to the computational basis (CB) by $b=\ddot{0}$. Thus we have
d+1 MUB bases,$ b=\ddot{0},0,1,...d-1$ with the total number of states d(d+1) grouped in
d+1 sets each of d states. We have of course,
\begin{equation}\label{mub}
\langle m;b|m';b\rangle=\delta_{m,m'};\;\;|\langle m;b|m';b'\rangle|=\frac{1}{\sqrt d},
\;\;b\ne b'.
\end{equation}
We mark states with double angular bracket ($|m;b\rangle\rangle$) to signify it to be an
eigenfunction of (the relevant) unitary operator $\omega^{b}\hat{X}\hat{Z}^{2b}$ for
b=0,1,...d-1 and of $\hat{Z}$ for $b=\ddot{0}$ (the latter signifies the computational
basis) and thus a an MUB state, cf. \cite{tal}.\\

\begin{acknowledgments}
I acknowledge with thanks numerous conversations and comments by my colleagues Professors
A. Mann and J. Slawny.
\end{acknowledgments}

\end{document}